\documentclass[aps,twocolumn,showpacs]{revtex4-1}

\usepackage{epsfig,graphics}

\def \E{{\rm e}}
\def \tr{{\rm Tr}\,}
\def \>{\rangle}
\def \<{\langle}

\newcommand\ket[1]{\ensuremath{|#1\rangle}}
\newcommand\bra[1]{\ensuremath{\langle#1|}}
\newcommand\iprod[2]{\ensuremath{\langle#1|#2\rangle}}
\newcommand\oprod[2]{\ensuremath{|#1\rangle\langle#2|}}

\begin{document}
\title{Optimal cloning of qubits given by
arbitrary axisymmetric distribution on Bloch sphere}

\author{Karol Bartkiewicz}

\affiliation{Faculty of Physics, Adam Mickiewicz University,
61-614 Pozna\'n, Poland}

\author{Adam Miranowicz}

\affiliation{Faculty of Physics, Adam Mickiewicz University,
61-614 Pozna\'n, Poland}

\begin{abstract}
We find an optimal quantum cloning machine, which clones qubits of
arbitrary symmetrical distribution around the Bloch vector with
the highest fidelity. The process is referred to as 
phase-independent cloning in contrast to the standard
phase-covariant cloning for which an input qubit state is {\em a
priori} better known. We assume that the information about the
input state is encoded in an arbitrary axisymmetric distribution
(phase function) on the Bloch sphere of the cloned qubits. We
find analytical expressions describing the optimal cloning
transformation and fidelity of the clones. As an illustration, we
analyze cloning of qubit state described by the von Mises-Fisher
and Brosseau distributions. Moreover, we show that the optimal
phase-independent cloning machine can be implemented by modifying
the mirror phase-covariant cloning machine for which quantum
circuits are known.

\end{abstract}
\date{\today}
\pagestyle{plain} \pagenumbering{arabic}

 \maketitle

%------------------------------------------------------------------
\section{Introduction}

The no-cloning theorem~\cite{Zurek} states that no quantum-mechanical
evolution exists which would transform an unknown quantum state $|\psi\>$
according to $|\psi\>\rightarrow |\psi\>|\psi\>$. This is provided by the linearity of
quantum mechanics. The no-cloning theorem guaranties, e.g., the
security (privacy) of quantum-communication protocols including
quantum key distribution.

Exact cloning is impossible; however, imperfect (optimal) cloning
is possible as it was first shown by Bu\v{z}ek and Hillery
\cite{Buzek96} by designing a $1\rightarrow 2$ cloning machine,
referred to as the {\em universal cloner} (UC). The cloning
machine prepares two approximate copies of an unknown pure qubit
state. The UC generates two qubit states with the same fidelity
$F=5/6$. Fidelities of the clones to the initial pure state are
equal $F_1=F_2$. Therefore, the UC is a state-independent
symmetric cloner.

It was later shown that for the $1\rightarrow M$ UC, a relation
between the optimum fidelity $F$ of each copy and the number $M$
of copies is given by $F=(2M+1)/(3M)$~\cite{Gisin}. Setting
$M\rightarrow\infty$ corresponds to a classical copying process
with $F=2/3$, which is the best fidelity that one can achieve with
only classical operations.

Further works have extended the concept to include cloning of
qudits, cloning of continuous-variable systems or state-dependent
cloning (non-universal cloning), which can produce clones of a
specific set of qubits with much higher fidelity than the
UC~\cite{Buzek98,Bruss98a,Niu,Bruss00,F1,F2, Demkowicz,
Fan,Buscemi,Sacchi,Kay,Hu,Bartkiewicz} (see also
reviews~\cite{cloning1,cloning2} and references therein).

The study of the state-dependent cloning machines is important because
it is often the case that we have some {\it a priori} information
on a given quantum state that we want to clone, but we do not
know it exactly. Then, by employing the available {\it a
priori} information, we can design a cloning machine which
outperforms the UC for some specific set of qubits.  For example,
if it is known that the qubit is chosen from the equator of the
Bloch sphere then the so-called {\em phase-covariant
cloners} (PCCs) have been designed~\cite{Bruss00,Fan}, and it was
shown to be optimal providing a higher fidelity than the UC.

Phase-covariant cloning of qubits has been further explored. For example,
Fiur\'a\v{s}ek~\cite{F2} studied the PCCs with known expectation
value of Pauli's $\sigma_z$ operator and provided two optimal
symmetric cloners: one for the states in the lower and the other
for those in the upper hemisphere of the Bloch sphere. Hu {\em et
al.}~\cite{Hu} studied phase-independent cloning of qubits uniformly
distributed on a belt of the Bloch sphere.
Bartkiewicz {\em et al.}~\cite{Bartkiewicz} provided an optimal
cloning transformation, referred to as the {\em mirror
phase-covariant cloning} (MPCC), for qubits of known modulus of
expectation value of Pauli's $\sigma_z$ operator.

Optimal cloning plays a crucial role in, e.g., quantum
cryptography. Security analyses of the quantum key distribution
protocols against coherent and incoherent attacks using quantum
cloners can be found in
Refs.~\cite{BB84,Fuchs97,Beckmann,Bruss98b,Imre}. Optimal
phase-independent cloners seem to play a special role there. One
example of the optimal phase-independent cloning machines is
the PCC, which can be used in an optimal attack on the BB84
quantum key distribution protocol~\cite{BB84,Fuchs97,Imre}.
Another example is the UC, which enables an optimal incoherent
attack on the six-state protocol~\cite{Beckmann,Bruss98b,Imre}.

Our paper is devoted to {\em phase-independent cloning},
which refers to cloning of qubits assuming that their
distribution (symmetrical around the Bloch vector) is {\em a
priori} known. It is related to the optimal state
estimation~\cite{Fuchs96} and to phase-independent telecloning and
telemapping~\cite{Kay}. Phase-independent cloning includes
the majority of all known optimal cloning machines as special
cases. One of the exceptions is the phase-dependent cloning
machine recently described in Ref.~\cite{Wu}.

In this paper, we find that phase-independent cloning
can be implemented analogously to the MPCC, e.g., in
linear-optical systems~\cite{Bartkiewicz2} or quantum
dots~\cite{Bartkiewicz} (see also Ref.~\cite{Zhu}).

Phase-independent cloning is an example of the optimal cloning
problem being invariant with respect to the discrete
Weyl-Heisenberg group (see, e.g., Ref.~\cite{cloning2}).

We also show here that the phase-independent cloning
exhibits sudden change in average single-copy fidelity, when the
symmetry of the system is reduced from $U(2)$ to $U(1)$.

Optimal cloning also limits the capacity of quantum channels. An
example is the Pauli channel and Pauli cloning machines analyzed
by Cerf~\cite{Cerf00}. Our results could be used in a
similar analysis for channels that undergo phase-independent
damping.

The paper is organized as follows. In Sec. II, we present a
general transformation describing an optimal symmetric $1\rightarrow
2$ cloning of a qubit. In Sec. III, we describe the optimal
symmetric $1\rightarrow 2$ cloning of a qubit knowing {\it a priori}
its axisymmetric distribution on the Bloch sphere. This cloning is
referred to as the optimal phase-independent cloning. In Sec. IV,
we analyze two examples of such cloning of qubits described by the
von Mises-Fisher and Brosseau distributions. In Sec. V, we
present, probably the most important result of the paper, the
optimality proof for the phase-independent cloning transformation.
In Sec. VI, we describe a quantum circuit implementing the
optimal phase-independent cloning. We conclude in Sec. VII.

%------------------------------------------------------------------
\section{Optimal symmetric cloning of qubits}

Suppose we want to clone a set of qubits, for which some
characteristic point is in the following {\em a priori}
known pure state:
\begin{equation}
\ket{\psi}=\cos{\frac{\vartheta}{2}}\ket{0}+\E^{i\varphi}\sin{\frac{\vartheta}
{2}}\ket{1}, \label{N01}
\end{equation}
which is parametrized by polar $\vartheta$ and azimuthal $\varphi$ angles
on the Bloch sphere. We can express all other qubit states as
\begin{equation}
\ket{\psi(\theta,\phi)} = \cos{\frac{\theta}{2}}\ket{\psi} +
\E^{i\phi}\sin{\frac{\theta}{2}}\ket{\bar{\psi}},\label{N02}
\end{equation}
where $\ket{\bar{\psi}}$ is a state orthogonal to
$\ket{\psi}$, i.e., $\iprod{\psi}{\bar{\psi}}=0$. Note that
$\cos{\theta}$ is equal to the scalar product of the Bloch vectors
$\vec{r}=[\langle\sigma_{x}\rangle,\langle\sigma_{y}
\rangle,\langle\sigma_{z} \rangle]$ and describes an elevation of
an arbitrary qubit with respect to the reference qubit $\ket{\psi}$,
whereas $\phi$ angle is an azimuth. Please note,
however, that $\theta$ is measured from $\ket{\psi}$.

The quality of the cloning can be described by the fidelity of a
single clone defined as
\begin{eqnarray}
F_{i}(\theta,\phi)&=& \bra{\psi(\theta,\phi)}
\rho_{i}\ket{\psi(\theta,\phi)},
 \label{N01a}
\end{eqnarray}
where $\rho_{i}=\tr_{i\oplus1}\left(\rho_{\rm out}\right)$ is the
reduced density matrix of the $i$th clone ($i=1,2$) and $\oplus$
denotes summation modulo 2. To obtain an optimal cloner
for an arbitrary distribution of qubits, we use average
single-copy fidelity $F$  in a way similar to that used for the
MPCC~\cite{Bartkiewicz2}. Thus, we express $F$ as
\begin{equation}
F=\frac12\int_0^{2\pi} \int_{-1}^{1} g(\theta,\phi)
\left[F_1(\theta,\phi)+F_2(\theta,\phi)\right]\, {\rm d}\cos\theta
{\rm d}\phi,\label{N01b}
\end{equation}
where $g(\theta,\phi)$ is an arbitrary distribution satisfying
the normalization condition
\begin{equation}
\int_{0}^{2\pi}\int_{-1}^{1} g(\theta,\phi)\,{\rm d}\cos{\theta}
{\rm d}\phi = 1.\label{N03}
\end{equation}
In the special case of 
\begin{equation}
g(\theta,\phi) =\frac1{4\pi\sin\vartheta}[\delta(\vartheta-\theta)
+\delta(\vartheta+\pi-\theta)], \label{N03a}
\end{equation}
our generalized cloner reduces to the MPCC.

The optimal cloning transformation can be found by maximizing
single-copy fidelity $F_{i}$, where $i=1$ or $2$, averaged over the distribution
$g(\theta,\phi)$ representing the frequency of occurrence of the
considered qubits~\cite{Bartkiewicz}.

Note that distributions $g$ describe our classical
knowledge about qubits to be cloned. Thus, the distributions $g$
are classical although cloning is quantum. Probably all works on
optimal quantum cloning are concerned with cloning of pure quantum
states. Indeed, a generic quantum cloning task for pure states can
be formulated such that the state to be cloned is drawn from a
certain ensemble of pure states with some {\em a priori}
probability and sent to the cloning machine which should produce
replicas of the input state. The quality of the cloner is then
usually characterized by mean clone fidelity averaged over the
whole ensemble. The probability distribution characterizing the
ensemble of states which one wants to clone is given by $g$ in our
case. So, $g$ is a classical probability distribution on a unit
sphere.

An optimal symmetric $1 \rightarrow 2$ cloning
transformation must be symmetric with respect to the swap
operation on clones. So, if we neglect the phase relations, the
optimal transformation for an arbitrary distribution of qubits can
be expressed as
\begin{eqnarray}
\ket{\psi}_{\rm in}\ket{00}_{1,2} &\rightarrow &
\cos\alpha_+(\cos{\beta_+}\ket{\psi,\psi}_{1,2}+\sin\beta_+\ket{\bar{\psi},\bar{
\psi}}_{1,2}) \nonumber \\&&\otimes\ket{\bar{\psi}}_3 +
\sin\alpha_+\ket{\Psi_+}_{1,2}\ket{\psi}_3,\nonumber \\
\ket{\bar{\psi}}_{\rm in}\ket{00}_{1,2} & \rightarrow &
\cos\alpha_-(\cos\beta_-\ket{\psi,\psi}_{1,2}+\sin\beta_-\ket{\bar{\psi},\bar{\psi}}_{1,2})
\nonumber \\&&\otimes\ket{\psi}_3   +
\sin\alpha_-\ket{\Psi_+}_{1,2}\ket{\bar{\psi}}_3,\label{N04}
\end{eqnarray}
where $\ket{\Psi_\pm}=
(\ket{\psi,\bar{\psi}}\pm\ket{\bar{\psi},\psi} )/\sqrt{2}$,
$\iprod{\bar{\psi}}{\psi}=0$ and subscripts 1 and 2 denote the
clones whereas 3 denotes the ancillary qubit. The angles
$\alpha_\pm$ and $\beta_\pm$ are found by maximizing the average
single-copy fidelity.

It is worth noting that we are {\em not} analyzing an optimal
cloning of mixed qubit states given by
\begin{equation}
\rho=\int_{0}^{2\pi}\int_{-1}^{1} g(\theta,\phi)
\ket{\psi(\theta,\phi)}\bra{\psi(\theta,\phi)}\, {\rm d}\cos{\theta} {\rm
d}\phi.
\end{equation}
By contrast, we are analyzing cloning of each of the pure qubit states
from an ensemble separately. While the distribution
$g(\theta,\phi)$ is a weight function for a single-copy fidelity
of a qubit state $\ket{\psi(\theta,\phi)}$.

%------------------------------------------------------------------
\section{Optimal cloning for axisymmetric distributions}

In the following, for simplicity, we consider only
axisymmetric distributions. The symmetry of the distributions
means that the cloning transformation is phase-independent (does
not depend on azimuthal angle $\phi$), i.e.,
distributions of qubits to be cloned are symmetrical along an
arbitrary $\ket{\psi}$ axis. Moreover, we first
postulate (as in Ref.~\cite{Hu}) that $\beta_\pm=0$ and then
(in Sec.~V) we prove that such transformation is optimal
for cloning.

When $\vartheta=\phi=0$, see Eq.~(\ref{N01}), all the formulas can
be rewritten in a computational basis by simply substituting
$\ket{\psi}$ and $\ket{\bar\psi}$ with $\ket{0}$  and $\ket{1}$,
respectively. Please note that one can proceed also in the
opposite direction; i.e., one can start assuming that the
distributions are centered around poles of the Bloch sphere (a
local coordinate system) and later transformed to the basis
so that the states match in, e.g., the laboratory frame (a
global coordinate system). We choose the first approach, since we
believe that it gives better physical intuition of the problem.

The density matrix of a single clone $\rho_i$ is derived by taking a
partial trace (over the ancillary qubit and one of the
clones) of the density matrix describing the system after the
cloning transformation. As a result we obtain:
\begin{eqnarray}
\rho_{i} &=& \frac{1}{2}\left[
\left((\cos^2{\alpha_+}+1)\cos^2{\frac{\theta}{2}}+
\sin^2{\alpha_-}\sin^2{\frac{\theta} {2}}\right)\oprod{\psi}{\psi}
\right.
\nonumber \\
&&+\left((\cos^2{\alpha_-}+1)\sin^2{\frac{\theta}{2}}+
\cos^2{\alpha_+}\cos^2{\frac{\theta}{2}}\right)\oprod{\bar{\psi}}{\bar{\psi}}
\nonumber \\
&&+\left. \left(
\frac{e^{-i\phi}}{\sqrt{2}}\sin{\theta}\sin{\Omega }
\oprod{\psi}{\bar{\psi}}  + {\rm h.c.} \right) \right],
\end{eqnarray}
where $i=1,2$ and $\Omega=\alpha_+ + \alpha_-$. Now, we can
express the single-copy fidelity, given by Eq.~(\ref{N01a}), as
\begin{eqnarray}
  F_i &=&
  \frac{1}{8} \left[2 \left(3 + \cos{2\alpha_+}\right) \cos^4{\frac{\theta}{2}}
+
   2 \left(3 + \cos 2 \alpha_-\right)\sin^4{\frac{\theta}{2}} \right.
   \nonumber \\
   &&+ \left.\left(\sin^2{\alpha_+} + \sin^2{\alpha_-} + 2 \sqrt{2}
\sin\Omega\right) \sin^2{\theta}\right].
\label{Fidelity}
\end{eqnarray}

We obtain expressions for $\alpha_\pm$ by maximizing, with respect
to $g(\theta)$, the average single-copy fidelity $F$, given by
Eq.~(\ref{N01b}). Any distribution axisymmetric function
$g(\theta)$ can be expressed in terms of the Legendre polynomials
$P_n(\cos{\theta})$ in the following way~\cite{Kaplan}:
\begin{eqnarray}
g(\theta) &=& \frac{1}{4\pi} \sum_{n=0}^{\infty} (2n+1)a_n P_n(\cos{\theta}),\\
a_n  &=& \int_{0}^{2\pi}\int_{-1}^{1}
g(\theta)P_n(\cos{\theta})\,{\rm d}\cos{\theta}\,{\rm d}\phi.
\label{an}
\end{eqnarray}
Thus, we present the results in terms of the expansion
coefficients $a_n$.

The form of the expressions for $\alpha_\pm$  depends on a single
parameter $\Gamma$, which is a generalization of the $T$ parameter
in Ref.~\cite{Hu}:
\begin{equation}
\Gamma = \frac{6\sqrt{2}a_1(a_2-1)}{x_+x_-},
\label{Gamma}
\end{equation}
where $x_\pm = 1+2a_2\pm3a_1$. We notice that for a normalized
qubit distribution ($a_0=1$) one needs to know only $a_1$ and
$a_2$ in order to fully characterize the corresponding optimal
cloning transformation. As long as $|\Gamma| < 1$, we can express
the parameters describing the cloning transformation as
\begin{eqnarray}
2\alpha_{\pm} =  \arcsin{\Omega} \pm \arcsin{\Gamma},
\label{alpha}
\end{eqnarray}
where
\begin{equation}
\Omega = \frac{2\sqrt{2}(1+2a_2)(1-a_2)}
{\sqrt{3x_+x_-(3+4a_2^2-3a_1^2-4a_2)}}. \label{Omega}
\end{equation}
However, if $|\Gamma|> 1$ then $\alpha_+ = 0$ and $\alpha_ -=
\frac{\pi}{2}$ or vice versa, and the optimal cloning
transformations corresponds to one of the transformation derived
by Fiur\'a\v{s}ek  in Ref.~\cite{F2}. The case of $a_1=0$ includes
the PCC for $\theta=\pi/2$ and the MPCC~\cite{Bartkiewicz}
($\alpha_+=\alpha_-$). Moreover, for $a_1=a_2=0$, we recover the UC
transformation~\cite{Buzek96}. Also the optimal cloning
transformation of Hu {\em et al.}~\cite{Hu} can be derived for an
arbitrary belt of the Bloch sphere.

%------------------------------------------------------------------
\section{Examples of phase-independent cloning}

\subsection{Optimal cloning of the von Mises-Fisher distribution}

%------------------------------------------------------------------
% fig.1
\begin{figure}[t!]
\hspace*{32mm}(a) \hspace*{32mm}(b)
 \epsfxsize=4.25cm\epsfbox{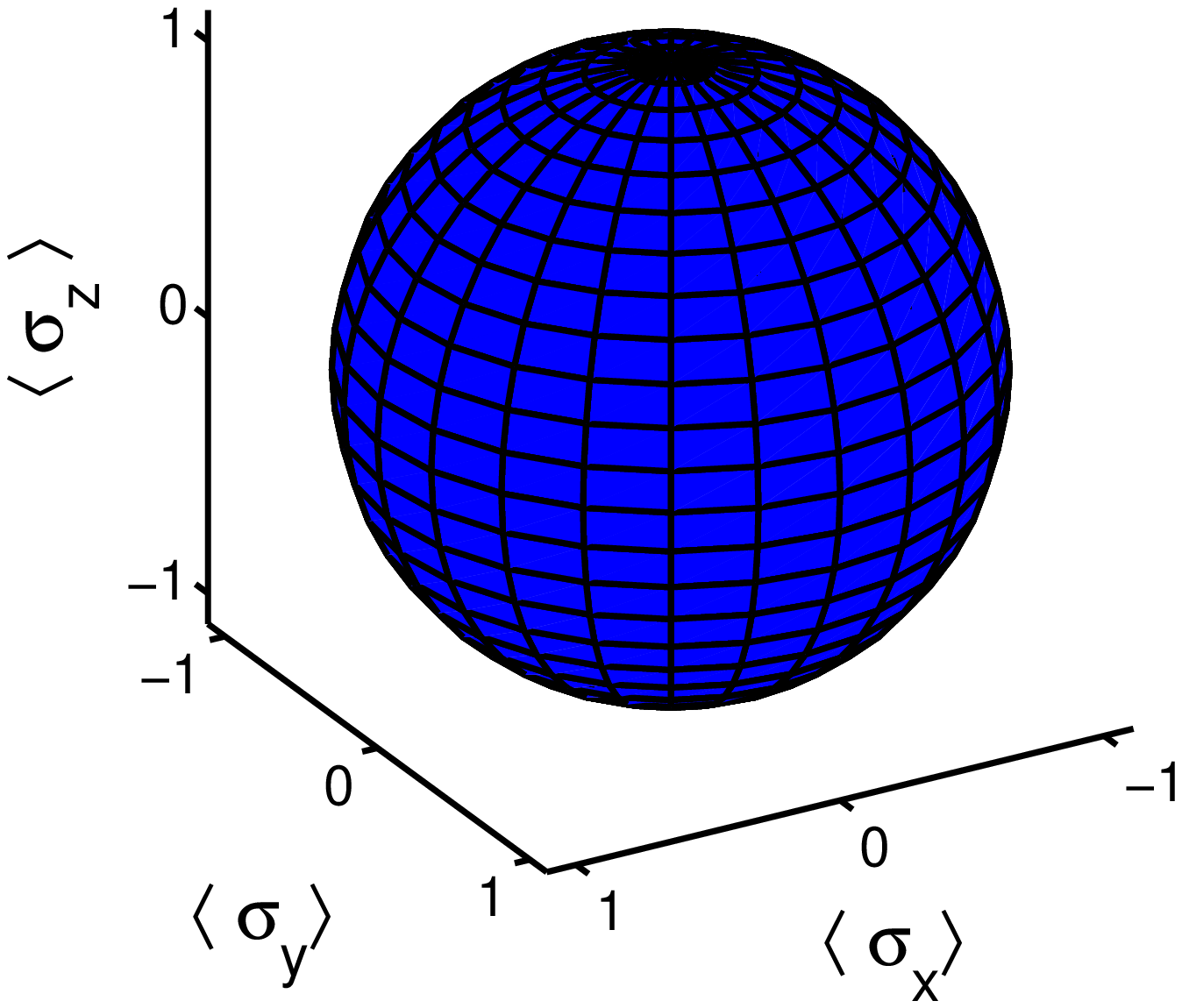}\hspace*{0mm}
 \epsfxsize=4.25cm\epsfbox{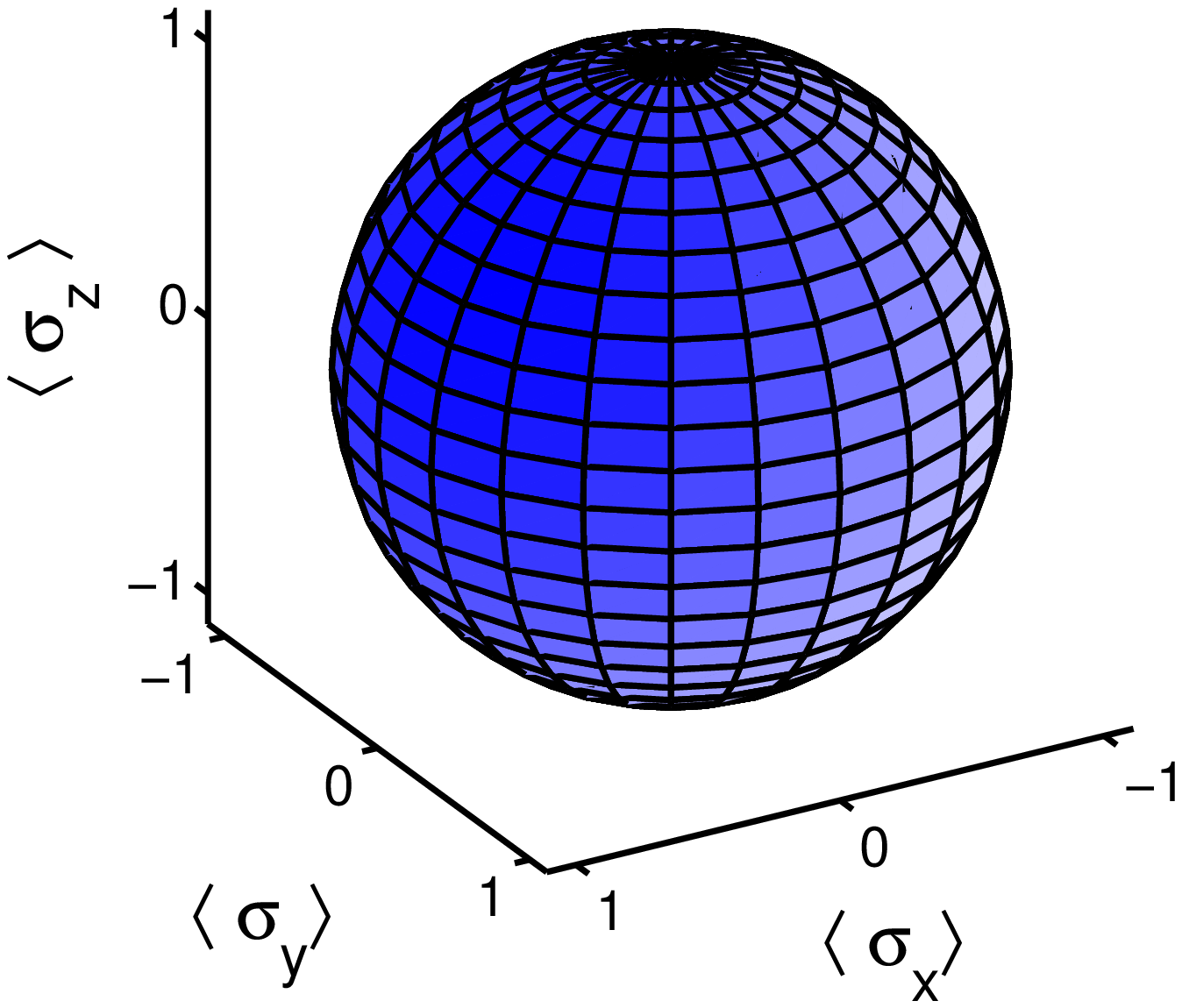}\vspace{0mm}\\
\hspace*{30mm}(c) \hspace*{32mm}(d)
 \epsfxsize=4.25cm\epsfbox{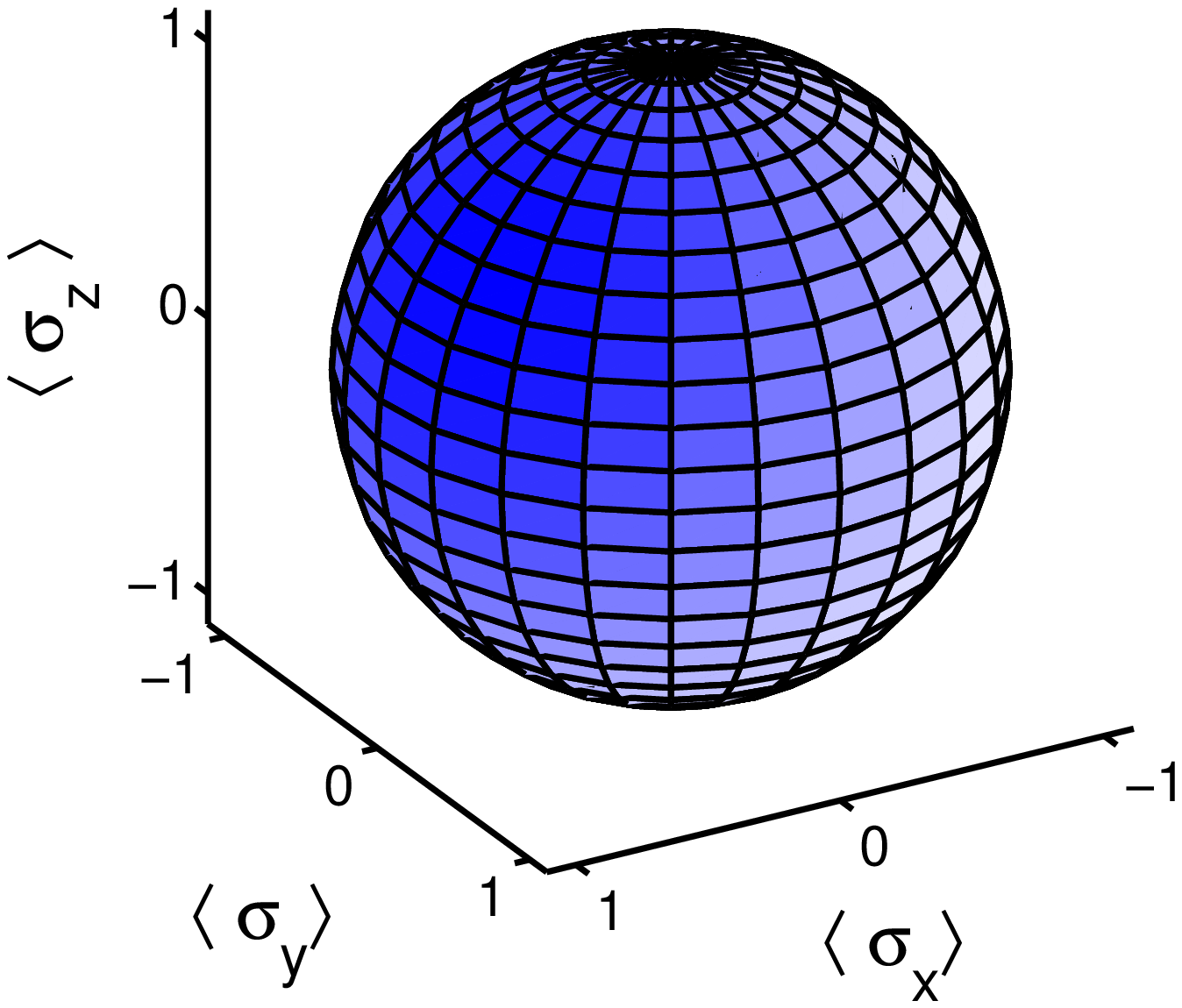}\hspace*{0mm}
 \epsfxsize=4.25cm\epsfbox{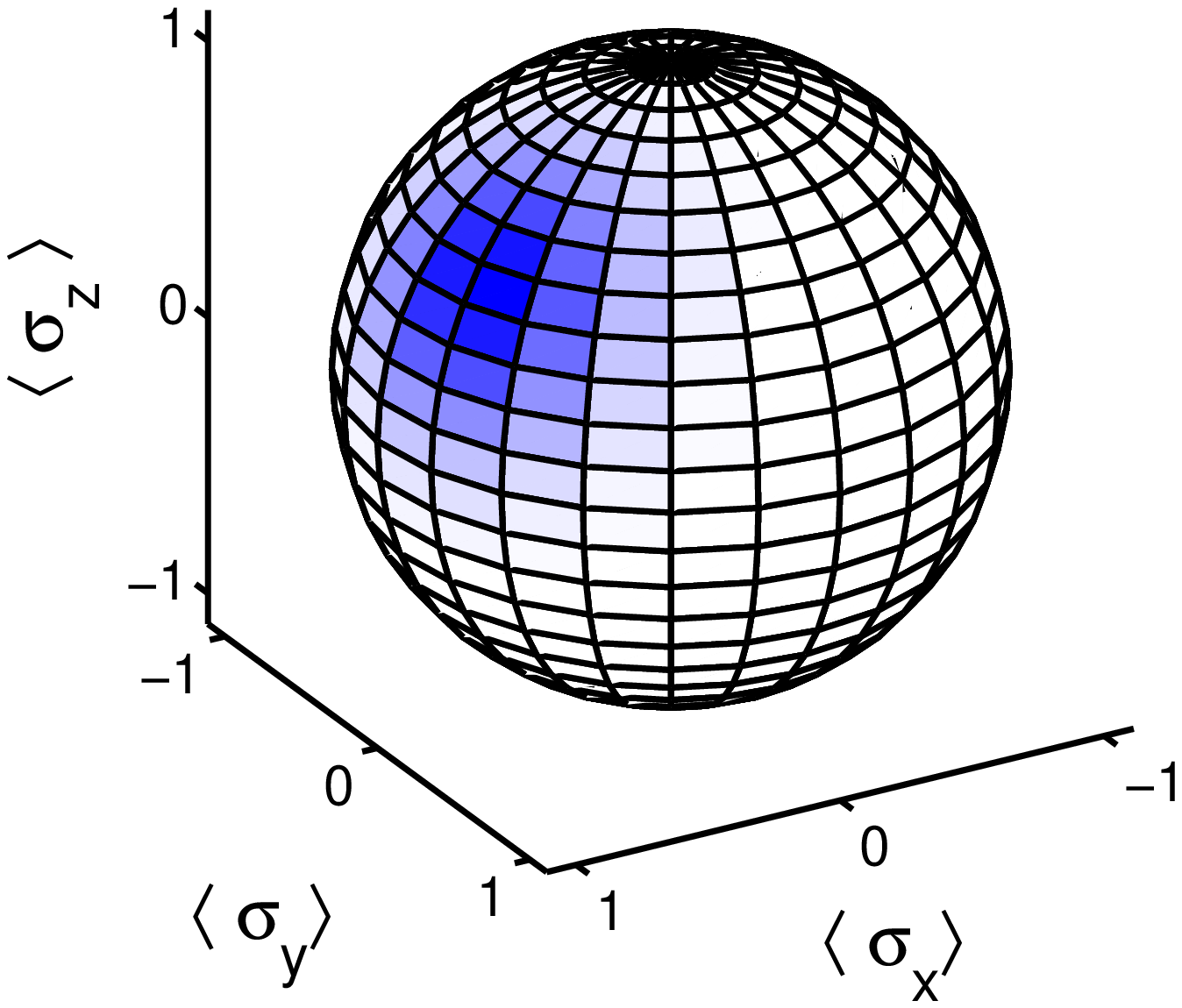}\vspace{0mm}
\caption[]{(Color online) Qubit states described on the Bloch
sphere by the von Mises-Fisher distribution (where the symmetry axis is rotated from the north pole
 by $\vartheta  = \pi/3$ and $\varphi = \pi/35$) divided by
its largest value for (a) $\kappa=0$, (b) $\kappa=0.3305$, (c)
$\kappa=1$ and (d) $\kappa=10$. The darker the region, the higher
the values of the probability distribution function. The mean
value $\langle \sigma_z \rangle = 1$. The optimal cloning
machines are the UC~\cite{Buzek96} for case (a) and the
PCC~\cite{F2} for the other cases. Note that the transition
area between (a) and (b) corresponds neither to the UC nor the
PCC.}\label{fig1}
\end{figure}

As an example of phase-independent cloning, let us analyze
the cloning of qubits described by the von Mises-Fisher distribution
(also called the Fisher distribution)~\cite{Fisher}, which is an
analog of the normal distribution but on a two-dimensional sphere
in ${\bf R}^{3}$ (see Fig.~\ref{fig1}). The distribution describes dispersion on a
sphere and has applications not only in physics and mathematical
statistics (especially directional statistics) but also in
quantitative biology, geology, or text data mining. It is a single-parameter
distribution which could be used to describe qubits
which undergo a random damping process; for example, weak
narrow-band light pulses propagating via real media or spin-$1/2$
systems in the presence of magnetic field fluctuations. The
distribution converges to a two-dimensional Gaussian
distribution for large values of $\kappa$. For $\kappa=0$ it
becomes a uniform distribution on a sphere.

\begin{figure}[t!]
\hspace{-3mm} \epsfxsize=9cm \epsfbox{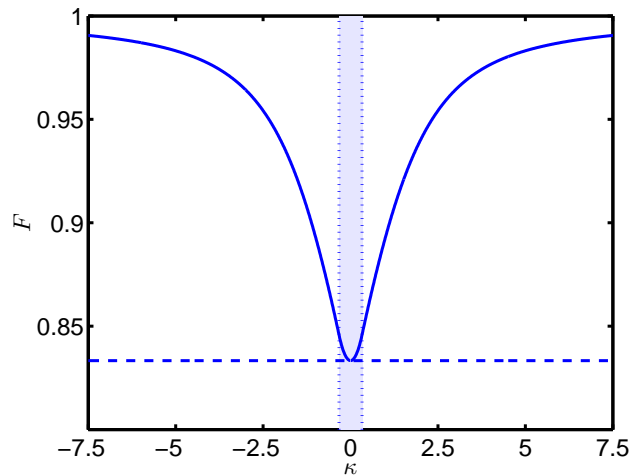}
 \caption[]{(Color online) Average single-copy fidelity $F$
vs concentration $\kappa$. The dashed line for $F=5/6$
corresponds to the UC limit, which is reached for $\kappa = 0$.
The shaded area corresponds to the range $-0.3305
<\kappa<0.3305$, where the PCC is no longer optimal.}\label{fig2}
\end{figure}

The von Mises-Fisher distribution is given by
\begin{equation}
g(\theta)=\frac{\kappa \exp{\left(\kappa
\cos{\theta}\right)}}{4\pi\sinh{\kappa}},
\end{equation}
where $\kappa$ is the concentration parameter (inverted variance),
which can also be negative when describing distribution
concentrated around $\ket{\bar{\psi}}$. For the von
Mises-Fisher distribution, we get the expansion coefficients,
defined in Eq.~(\ref{an}), as follows:
\begin{eqnarray}
a_1&=&-1/\kappa + \coth\kappa, \nonumber\\
a_2&=&1-3\kappa a_1. \label{aF}
\end{eqnarray}
If $\kappa \geq 0.3305$ then the PCC transformation of
Fiur\'a\v{s}ek's~\cite{F2} is optimal. On the other hand, if
$\kappa=0$ the optimal cloning transformation corresponds to the
UC~\cite{Buzek96}. So, practically, a very small axisymmetric
concentration of qubits (small value of $\kappa$), as shown in
Fig.~\ref{fig1} (Fig.~\ref{fig2}), is enough to determine which of
the two cloning transformations is optimal.

%------------------------------------------------------------------
% fig.2

We can see in Fig.~\ref{fig2} that the optimal cloning is related to the symmetry of
the cloned set of qubits (see, e.g., Ref. \cite{Chiribella}). As
one would expect, there is a sudden change in the average
single-copy fidelity when the set of cloned qubits reduces its
symmetry from $U(2)$ to $U(1)$.

It is somewhat surprising that the von Mises-Fisher distribution,
which is an analog of Gaussian distribution on a sphere, is not
widely used in Monte Carlo simulations. This is because of its
computational overhead. A more popular one-parameter phase
distribution on a sphere is the Henyey-Greenstein
function~\cite{Henyey}, for which $a_n = h^n$ and $h$ is the
anisotropy factor equal to the average of $\cos\theta$. The
Henyey-Greenstein function is widely used in simulations of, e.g.,
volume scattering processes. The explicit form of this function is
similar to the Brosseau distribution discussed in the following
subsection.

%------------------------------------------------------------------
\subsection{Optimal cloning of the Brosseau distribution}

%------------------------------------------------------------------
% fig.3
\begin{figure}[t!]
\hspace{-3mm} \epsfxsize=9cm \epsfbox{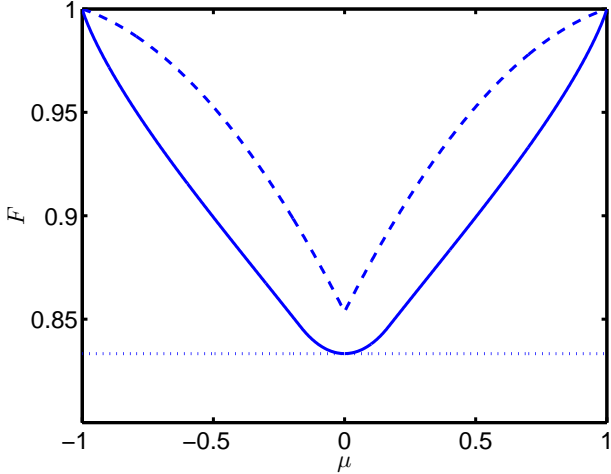}
 \caption[]{(Color online) Average single-clone
fidelity $F$ vs parameter $\mu$ for $\mu^2=P^2$ (solid) and
$\mu\neq P=1$ (dashed curve). The dotted line for $F=5/6$ shows
the UC limit, which is reached for $\mu = P=0$, whereas the dashed
curve corresponds to the fidelity of the PCC.}\label{fig3}
\end{figure}

%------------------------------------------------------------------
% fig.4
\begin{figure}[t!]
\hspace*{32mm}(a) \hspace*{32mm}(b)
 \epsfxsize=4.25cm\epsfbox{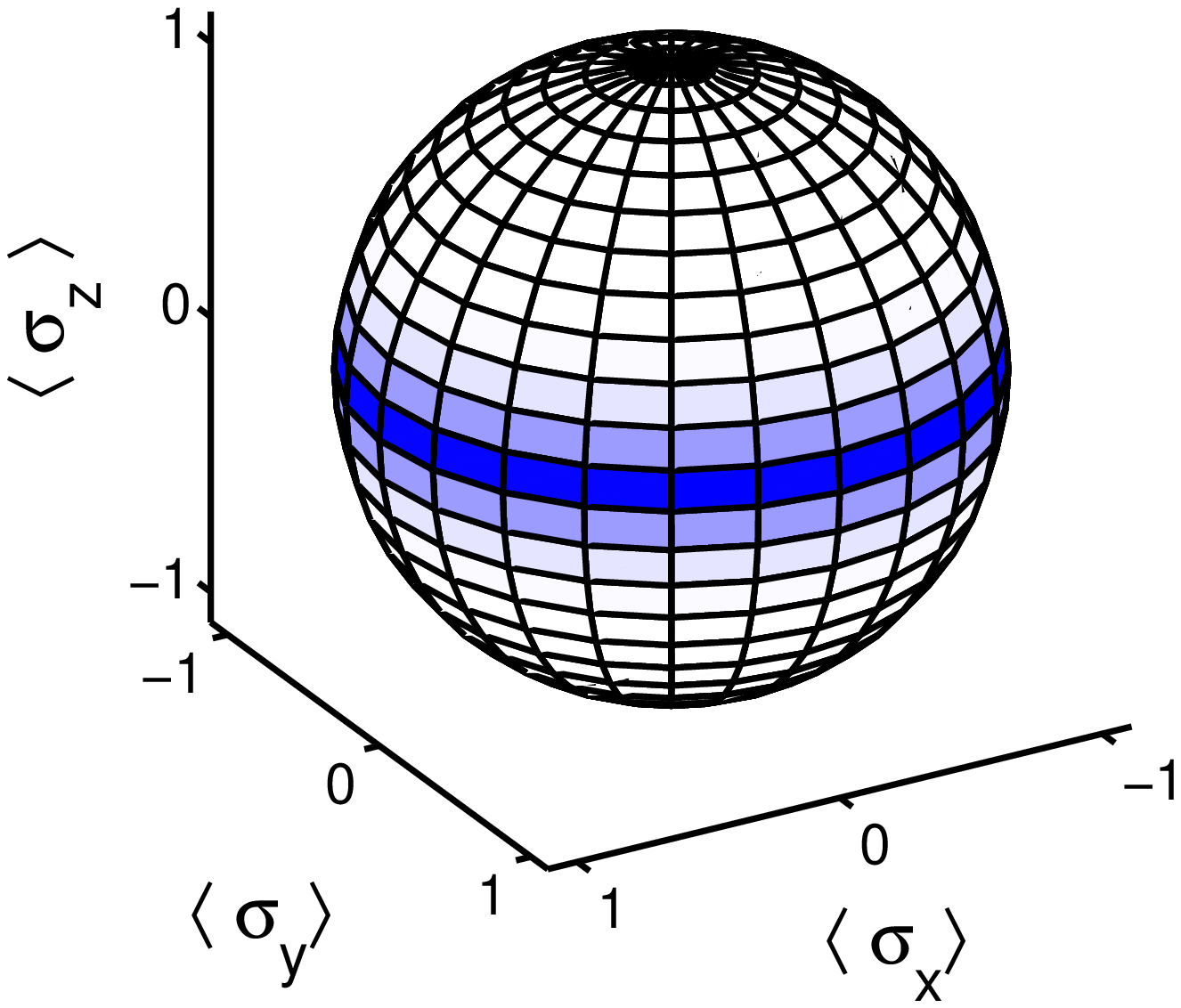}\hspace*{0mm}
 \epsfxsize=4.25cm\epsfbox{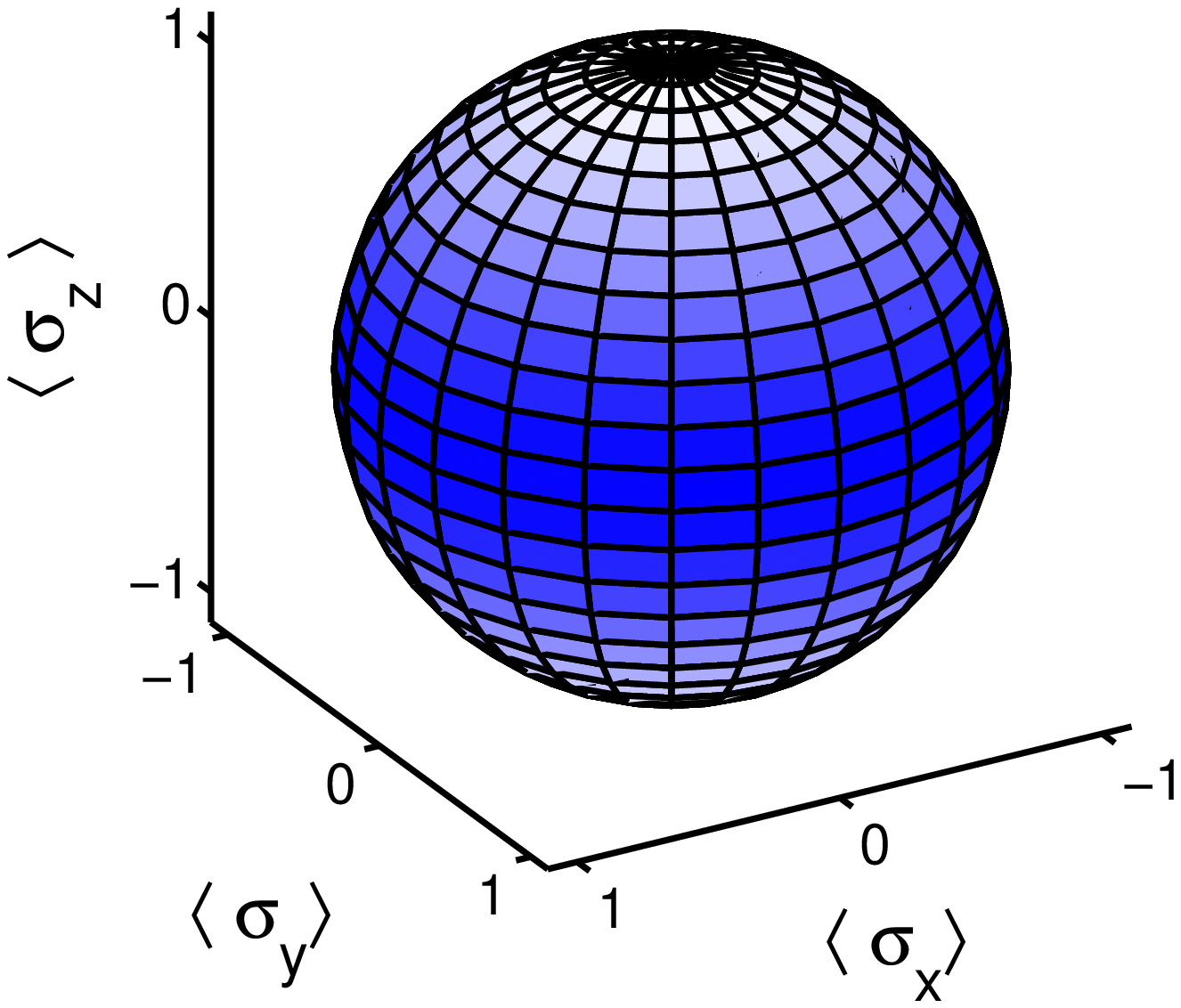}\vspace{0mm}\\
\hspace*{30mm}(c) \hspace*{32mm}(d)
 \epsfxsize=4.25cm\epsfbox{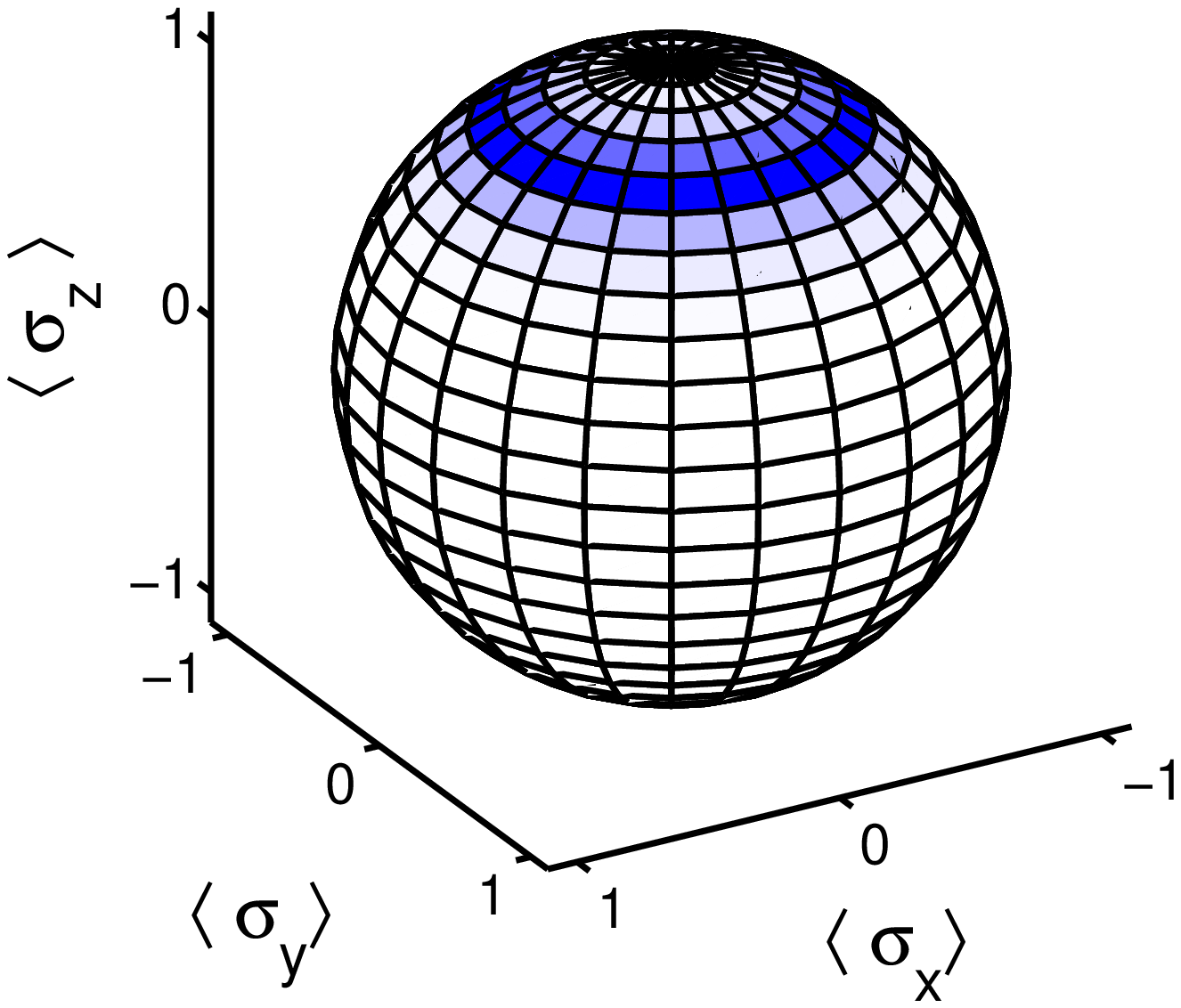}\hspace*{0mm}
 \epsfxsize=4.25cm\epsfbox{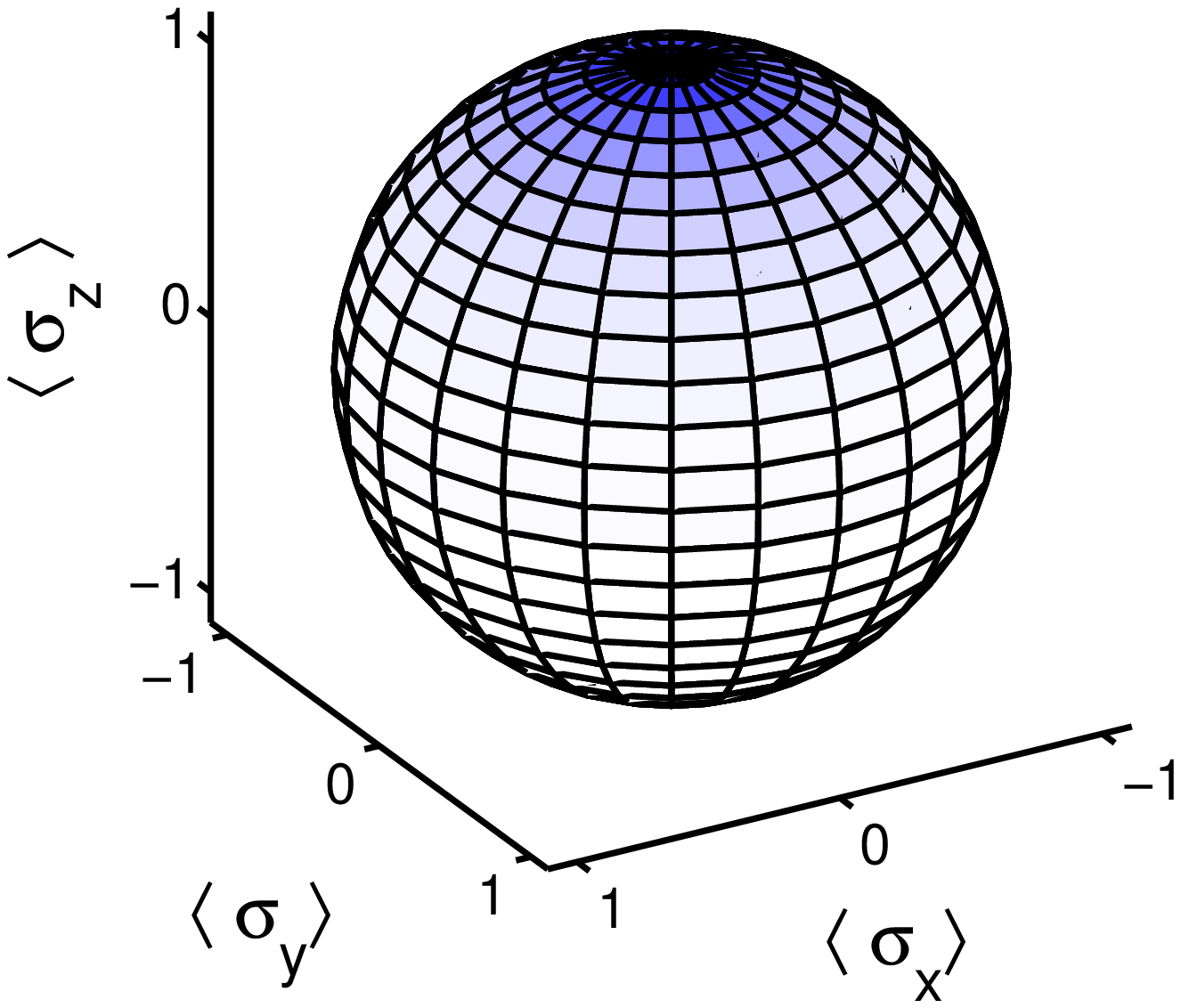}\vspace{0mm}
\caption[]{(Color online) Same as Fig.~1, but for the Brosseau
distribution assuming (a) $\mu=0$, $P=0.99$, (b) $\mu=0$,
$P=0.65$, (c) $\mu=0.8$, $P=0.99$, and (d) $\mu=0.8$, $P=0.8$. The
optimal cloning machines are the MPCC~\cite{Bartkiewicz} with
$\Lambda = \cos\alpha_+$ for case (a) and the PCC~\cite{F2} for
the other cases.}\label{fig4}
\end{figure}

Another example of phase-independent cloning is cloning of photons
of known statistics of a single Stokes parameter, e.g., $S_1$.
Statistics of the Stokes parameters was studied in detail by
Barakat~\cite{Barakat}; however, for us the statistics of
the normalized Stokes parameters is of greater interest since it
describes the state of a single photon.

In the quantum regime, the normalized Stokes parameters $s_i=S_i/S_0$
for $i=1,2,3$ correspond to Pauli's operators in the following way:
$s_1 = \sigma_z$, $s_2 =\sigma_x$, and $s_3 = \sigma_y$. Here, the
matrices are defined in the basis of $\{\ket{\psi}=\ket{H},
\ket{\bar{\psi}}=\ket{V}\}$, i.e., for the horizontal and vertical
polarization states, correspondingly.

For a Gaussian stochastic plane-wave field, the normalized Stokes
$s_1$ parameter has the following probability distribution
function derived by Brosseau~\cite{Brosseau}:
\begin{equation}
g(\theta) = \frac{(1-P^2)(1-\mu\cos\theta)}{2[
(1-\mu\cos\theta)^2-(1-\cos^2\theta)(P^2-\mu^2)]^{\frac{3}{2}}}.
\label{N14}
\end{equation}
where $\cos\theta=s_1$ is the Stokes parameter, $P$ is the degree
of polarization, and $\mu=\langle S_1\rangle/\langle S_0\rangle$;
moreover, $P^2-\mu^2 \geq 0$. The Brosseau distribution has
the same asymptotic behavior as the von Mises-Fisher distribution
if analyzed for two limiting values of $P$. For $P=0$, we have
unpolarized light and the resulting distribution is uniform, which
corresponds to the UC. For $P=1$, the polarization can be described
as $\cos{\theta}\ket{H} +\exp(i\phi)\sin{\theta}\ket{V}$, where
$\theta$ is fixed and $\phi$ is an unknown constant. So, the
distribution converges to Dirac's $\delta$ function
$\delta(\cos\theta-\mu)$, which corresponds to the PCC~\cite{F2}.
Then the distribution is given by two physical parameters $P$ and
$\mu$ describing our knowledge about the polarization state of the
photon. The lowest degree of polarization is given as $P^2=\mu^2$,
this corresponds to a situation, in which phase $\phi$ is random
(see Fig.~\ref{fig3}). By varying those two parameters, we can
construct various qubit distributions, as shown, e.g., in
Fig.~\ref{fig4}. For the Brosseau~\cite{Brosseau}
distribution, the $a_i$ coefficients are more than those in
Eq.~(\ref{aF}) as given by:
\begin{eqnarray}
a_1&=& \frac{1-P^2}{2}(I_1-\mu I_2), \nonumber\\
a_2&=& \frac{3}{4}(1-P^2)(I_2-\mu I_3)-\frac{1}{2},
\label{aB}
\end{eqnarray}
where
\begin{eqnarray}
I_n=\int_{-1}^{1}\frac{x^n}{(1+\mu^2-P^2-2x\mu+x^2P^2)^{3/2}}
\,{\rm d}x.
\end{eqnarray}

It is worth noting that the field discussed by Brosseau
in Ref.~\cite{Brosseau} was analyzed using the classical
description of the polarized light. Namely, he considered a
Gaussian stochastic plane-wave field. In his main formula,
Brosseau used the following notations: the angular brackets denote
the time average, while $P$ is the classical degree of
polarization. So, we use Eq.~(\ref{N14}), which was obtained using
a classical description of light for a quantized field. Thus, one
can raise an objection that the use of a final classical
distribution $g(\theta)$, obtained using the classical degree of
polarization and the time average of the Stokes parameters, in the
case of a quantum description of the field is incorrect.

To show that the application of classical distributions
$g(\theta)$ for quantum cloning is indeed correct, let us clarify
the following point: The optimal quantum phase-cloning described
in the present manuscript and in the vast majority of papers on
quantum cloning (see reviews~\cite{cloning1,cloning2} and
references therein) is {\em not} cloning of mixed qubit states,
but it is cloning of a pure qubit state from an ensemble. We know
{\em a priori} that the ensemble is described by some distribution
$g(\theta)$ which represents our classical partial knowledge about
the qubit to be cloned. Thus, the distribution is classical
although cloning transformation is quantum.

Our application of classical distribution $g(\theta)$ is in
complete agreement with other works on optimal cloners (see Table~\ref{summary}). Our
examples include (i) the universal cloners of Bu\v{z}ek and
Hillery~\cite{Buzek96}, where the classical distribution
$g(\theta)$ is equal to $\frac1{4\pi}$, (ii) the optimal
phase-covariant cloners of Bru\ss{} {\em et
al.}~\cite{Bruss98a,Bruss00} (for $\vartheta=\pi/2$) and of
Fiur\'a\v{s}ek~\cite{F1,F2} (for any $\vartheta$), where
$g(\theta)$ is given by Dirac's $\delta$ function $\frac
1{2\pi\sin(\vartheta)}\delta(\vartheta-\theta)$, (iii) the optimal mirror
phase-covariant cloners of Bartkiewicz {\em et
al.}~\cite{Bartkiewicz}, where $g(\theta)$ is given by Eq.~(6),
and (iv) the optimal quantum cloners of a state in a belt of Bloch
sphere as described by Hu {\em et al.}~\cite{Hu}, where
$g(\theta)$ is given by a rectangle distribution, i.e., states are
assumed to be uniformly distributed in a region forming a belt
between two latitudes on the Bloch sphere.

\begin{table*}[!ht]
\caption{Summary of optimal phase independent symmetric $1\rightarrow 2$ cloning machines for qubits.}

\vspace{.3cm}
\begin{ruledtabular}
\begin{tabular}{ccc}
Optimal cloning machine & Distribution $g(\theta)$ & Single-copy fidelity $F$\\
\hline 
Universal cloner    (UC)        & \begin{large} $\frac1{4\pi}$                                                                                       \end{large} & 
\begin{large} $\frac{5}{6}$ \end{large}\\[5pt]
Phase-covariant cloner (PCC)\footnotemark[1]       & \begin{large} $\frac{\delta(\vartheta-\theta)}{2\pi\sin\vartheta}$             \end{large}                                   &
 $\frac18\left[5+\sqrt{2}+2\cos(\theta+\kappa)-(\sqrt{2}-1)\cos(2\theta)\right]$\\[5pt]
Mirror phase-covariant cloner (MPCC)\footnotemark[2]      & \begin{large} $\frac{\delta(\vartheta-\theta)+\delta(\vartheta+\pi-\theta)}{4\pi\sin\vartheta}$  \end{large}                 & 
$\frac{1+\Lambda^2}{2}-\frac{1}{2}\sin^2\theta\left(\Lambda^2-\Lambda\bar{\Lambda}\sqrt{2}\right)$\\[5pt]
Cloner of Bloch-sphere belt\footnotemark[3]$^{,}$\footnotemark[4]      & \begin{large} $\frac{u(\theta-\vartheta_1) -u(\theta-\vartheta_2)}{2\pi(\cos\vartheta_1 -\cos\vartheta_2)}$ \end{large}      &
Eq.~(\ref{Fidelity}) with Eq.~(15)\footnotemark[5] form Ref.~\cite{Hu} \\[5pt]
Cloner of von Mises-Fisher distribution\footnotemark[3] & \begin{large} $\frac{\kappa \exp{\left(\kappa\cos{\theta}\right)}}{4\pi\sinh{\kappa}}$                   \end{large}         &
Eq.~(\ref{Fidelity}) with Eq.~(\ref{aF}) \\[5pt]
Cloner of Brosseau distribution\footnotemark[3]  & \begin{large} $\frac{(1-P^2)(1-\mu\cos\theta)}{2[(1-\mu\cos\theta)^2-(1-\cos^2\theta)(P^2-\mu^2)]^{\frac{3}{2}}}$\end{large} &
Eq.~(\ref{Fidelity}) with Eq.~(\ref{aB})\\[5pt]
\end{tabular}
\end{ruledtabular}
\footnotetext[1]{$\kappa=0$ for $0\le\vartheta<\frac{\pi}2$ and $\kappa=\pi$ for $\frac{\pi}2\le\vartheta\le\pi$.}
\footnotetext[2]{$\Lambda=\sqrt{1-\bar{\Lambda}^2}$, $\Lambda$ depends on $\vartheta$ as given in Ref.~\cite{Bartkiewicz}.}
\footnotetext[3]{The explicit expression for $F$ is lengthy and therefore is not shown here, but it can be directly derived form the quoted equations.}
\footnotetext[4]{$u$ is Heviside's step function.}
\footnotetext[5]{In Ref.~\cite{Hu} $\alpha\equiv\alpha_+$ and $\beta\equiv\alpha_-$.}
\label{summary}
\end{table*}

%------------------------------------------------------------------
\section{Optimality proof for the phase-independent cloning transformation}

Cloning transformations can be described by a completely
positive trace-preserving map (CPTP) $\chi$, which can be written
with the use of the inverse  Jamio\l{}kowski isomorphism
\cite{Jamiolkowski} as follows:
\begin{equation}
\rho_{\rm out}=\tr_{\rm in}(\chi\rho_{\rm in}^T\otimes\openone^{\otimes
2}).
\end{equation}
The average single-copy fidelity (the figure of merit when
maximized) can be expressed as
\begin{equation}
F=\tr(\chi R),
\end{equation}
and
\begin{equation}
R =\frac{1}{2}\int_{0}^{2\pi}\int_{-1}^{1} \rho_{\rm
in}^T\otimes\left(\rho_{\rm in}\otimes\openone + \openone\otimes\rho_{\rm in}\right)
g(\theta)\,{\rm d}\cos{\theta} \,{\rm d}\phi.
\end{equation}
Since the cloning transformation for the set of cloned qubits
needs to be symmetric with respect to an arbitrary rotation $U$
about the state $\ket{\psi}$ and to swapping $U_{\rm SWAP}$
of the clones, the following commutation relations should
hold:
\begin{eqnarray}
\left[ \chi,U^{*}\otimes U^{\otimes 2} \right] &=&0,  \nonumber \\
\left[ \chi, \openone\otimes {\rm U_{\rm SWAP}}\right] &=&0.
\end{eqnarray}
The commutation relations and trace-preserving condition
restrict the $\chi$ matrix to the following form (in
$\{\ket{\psi},\ket{\bar{\psi}}\}^{\otimes 3}$ basis):
\begin{eqnarray}
\chi & =& \left[ \begin{array}{cccccccc}
\eta_1 & 0 & 0 & 0 & 0& \zeta_1& \zeta_1 & 0 \\
 0 & \eta_2 & \eta_3 & 0 & 0 & 0 & 0 & \zeta_2\\
 0 & \eta_3 & \eta_2 & 0 & 0 & 0 & 0 & \zeta_2\\
 0 & 0 & 0 & \eta_4 & 0 & 0 & 0 & 0\\
 0 & 0 & 0 & 0 &\xi_4   & 0 & 0 & 0 \\
\zeta_1 & 0 & 0 & 0 & 0 & \xi_2 & \xi_3 & 0 \\
\zeta_1 & 0 & 0 & 0 & 0 &  \xi_3 & \xi_2& 0 \\
 0 & \zeta_2 & \zeta_2 & 0 & 0 & 0 & 0 & \xi_1 \end{array} \right],
\end{eqnarray}
where $\xi_4=1-2\xi_2-\xi_1$ and $\eta_4=1-2\eta_2-\eta_1$. It is
more convenient to use another basis to express $\chi$ and
$R$ operators, i.e.: $\{ \ket{\psi,\psi,\psi},
\ket{\bar{\psi}}\ket{\Psi_+}, \ket{\bar{\psi},\bar{\psi},
\bar{\psi}}, \ket{\psi}\ket{\Psi_+},$ $
\ket{\bar{\psi}}\ket{\Psi_-}, \ket{\psi}\ket{\Psi_-},
\ket{\psi,\bar{\psi},\bar{\psi}}, \ket{\bar{\psi},\psi,\psi}\}$.
In this new basis the operators have the following block form:
\begin{eqnarray}
\chi =  \bigoplus_{i=1}^{6} \chi_i, \quad\quad R =
\bigoplus_{i=1}^{6} R_i,
\end{eqnarray}
where $\chi_{1(2)}$ and $R_{1(2)}$ are $2\times 2$ matrices. The
$\chi_i$ and $R_i$ elements for $i=3,...,6$ are not negative
numbers limited by trace of $\chi$ ($\tr\chi=2$). The average
single-copy fidelity can be expressed using Einstein's summation
convention as
\begin{equation}
F = \tr{(R_i\chi_i)}.
\end{equation}

The fidelity is a convex superposition and the matrix elements
$\chi_i$ (for $i=3,4,5,$ and $6$) are numbers $\in [0,2]$.
Therefore, it can be proved that for any distribution
$g(\theta)$ for $i\in\{5,6\}$ we have $\tr{(R_i\chi_i)} \leq 1/2$.
Moreover, for $i\in\{3,4\}$ we have $\tr{(R_i\chi_i)} \leq F_{\rm PCC}$,
where $F_{\rm PCC}$ is the single-copy fidelity of
the PCC averaged over inputs characterized by $g(\theta)$.
The PCC is derived only from $\chi_1$ and $\chi_2$.

Hence, in order to maximize
the fidelity we have to put $\chi_3=\chi_4=\chi_5=\chi_6=0$. This
implies that $\eta_2=\eta_3$, $\xi_2=\xi_3$ and
$1-\eta_1=2\eta_2$, $1-\xi_1=2\xi_2$. Now, the CPTP map depends
only on four parameters and can be described as a direct sum
(denoted by $\oplus$) of two matrices:
\begin{eqnarray}
\chi  = \left[ \begin{array}{cc}
\eta_1 & \sqrt{2}\zeta_1 \\
\sqrt{2}\zeta_1 & 1-\xi_1 \end{array} \right] \oplus \left[
\begin{array}{cc}
\xi_1 & \sqrt{2}\zeta_2\\
\sqrt{2}\zeta_2 & 1-\eta_1 \end{array} \right],
\end{eqnarray}
where the first matrix acts in a subspace spanned by
$\{\ket{\psi,\psi,\psi}, \ket{\bar{\psi},\Psi_+} \}$ and the
second by $\{\ket{\bar{\psi},\bar{\psi},\bar{\psi}},
\ket{\psi,\Psi_+}\}$. The map can be parametrized without the
loss of generality in the following way:
\begin{equation}
\chi  = \left[ \begin{array}{cc}
\cos^2\alpha_+ & \sqrt{2}\zeta_1 \\
\sqrt{2}\zeta_1 &  \sin^2\alpha_- \end{array} \right] \oplus
\left[ \begin{array}{cc}
\cos^2\alpha_- & \sqrt{2}\zeta_2\\
\sqrt{2}\zeta_2 & \sin^2\alpha_+ \end{array} \right] ,
\end{equation}
However, for the extremal CPTP maps~\cite{Chiribella}, we have that
$\chi_{1(2)}^2\propto\chi_{1(2)}$. From this condition, we find
 that $\sqrt{2}\zeta_1  = \cos\alpha_+ \sin\alpha_-$ and
$\sqrt{2}\zeta_2 = \cos\alpha_- \sin\alpha_+$. Finally, any CPTP
map that maximizes the average single-copy fidelity for an arbitrary
axisymmetric distribution $g(\theta)$ can be written  in the
basis $\{\ket{\psi},\ket{\bar{\psi}}\}^{\otimes 3}$  as
\begin{eqnarray}
\chi  = \left[ \begin{array}{cccccccc}
c^2_+ & 0 & 0 & 0 & 0& \frac{s_-c_+}{\sqrt{2}}& \frac{s_-c_+}{\sqrt{2}} & 0 \\
 0 & \frac{s^2_+}{2} & \frac{s^2_+}{2} & 0 & 0 & 0 & 0 &
\frac{c_-s_+}{\sqrt{2}}\\
 0 & \frac{s^2_+}{2} & \frac{s^2_+}{2} & 0 & 0 & 0 & 0 &
\frac{c_-s_+}{\sqrt{2}}\\
 0 & 0 & 0 & 0 & 0 & 0 & 0 & 0\\
 0 & 0 & 0 & 0 & 0 & 0 & 0 & 0 \\
\frac{s_-c_+}{\sqrt{2}} & 0 & 0 & 0 & 0 & \frac{s^2_-}{2} &
\frac{s^2_-}{2} & 0
\\
\frac{s_-c_+}{\sqrt{2}} & 0 & 0 & 0 & 0 &  \frac{s^2_+}{2} &
\frac{s^2_-}{2}& 0
\\
 0 & \frac{c_-s_+}{\sqrt{2}} & \frac{c_-s_+}{\sqrt{2}} & 0 & 0 & 0 & 0 & c^2_-
\end{array} \right]\!\!,
\end{eqnarray}
where $c_\pm=\cos\alpha_\pm$ and $s_\pm=\sin\alpha_\pm$. The above
map can be decomposed into unitary transformations, given by
Eq.~(\ref{N04}) (for $\beta_\pm=0$), by means of the Kraus decomposition~\cite{Kraus}.
This completes the proof.

%------------------------------------------------------------------
\section{Quantum circuit for optimal phase-independent cloning}

\begin{figure}[t!]
\hspace{-3mm} \epsfxsize=8cm \epsfbox{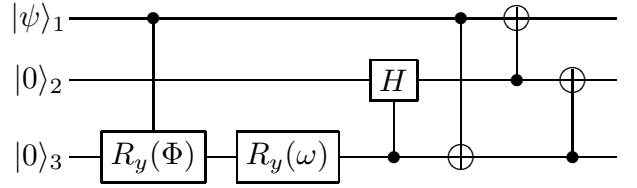} \caption[]{A
quantum circuit which implements the optimal phase-independent
cloning transformation of $\ket{\psi(\theta,\phi)}=\cos{(\theta/2)}\ket{0}
+\exp{(i\phi)}\sin{(\theta/2)}\ket{1}$ described on the Bloch
sphere by arbitrary axisymmetric distribution function
$g(\theta)$. From left to right: controlled rotation $R_y$ about
the $y$ axis by angle  $\Phi=2(\alpha_--\alpha_+)$, rotation $R_{y}$
by angle  $\omega=2\alpha_+$, controlled Hadamard gate, and three
CNOT gates.}\label{fig5}
\end{figure}

The analyzed cloning problem can be also expressed in the logical
basis. The optimal cloning transformation can be now written as
\begin{eqnarray}
\ket{000} &\rightarrow & \cos\alpha_+\ket{001} +
\sin\alpha_+\ket{\psi_+}\ket{0},
 \nonumber \\
\ket{100} & \rightarrow & \cos{\alpha_-}\ket{110} +
\sin{\alpha_-}\ket{\psi_+}\ket{1},
\end{eqnarray}
where $\ket{\psi_+}=\left(\ket{01} + \ket{10} \right)/\sqrt{2}$.
The quantum circuit, shown in Fig.~\ref{fig5}, performs the
following transformation:
\begin{equation}
|\psi_{\rm out}\rangle =
 U_{\rm CNOT}^{(32)}U_{\rm CNOT}^{(21)}U_{\rm CNOT}^{(13)}
 U^{(32)}_{\rm CH}U_{\rm R_y}^{(3)}U_{\rm CR_y}^{(13)}|\psi_{\rm in}\rangle,
 \label{N101b}
\end{equation}
where the superscripts indicate qubits for which the corresponding
gate is applied. The basic elements of the circuit are the
rotation
\begin{eqnarray}
U_{\rm R_y}(\omega)&=& \left[\begin{array}{cc}
\cos{(\omega/2)} & -\sin{(\omega/2)}\\
\sin{(\omega/2)} &  \cos{(\omega/2)}\\
\end{array}\right]
\label{N101c}
\end{eqnarray}
about the $y$ axis by angle $\omega=2\alpha_+$ and the controlled
rotation $U_{\rm CR_y}(\Phi)$ by angle
$\Phi=2(\alpha_--\alpha_+)$. In addition, this circuit is composed
of the controlled NOT (CNOT) gates, $U_{\rm CNOT}$, and the
controlled Hadamard gate, $U_{\rm CH}$, which can be
decomposed~\cite{Bartkiewicz} as $U^{(32)}_{\rm
CH}=A^{(2)}U^{(32)}_{\rm CNOT}A^{(2)}$, where
\begin{equation}
A=\frac{1}{\sqrt{4+2\sqrt{2}}}\left[\begin{array}{cc}
1 & 1+\sqrt{2}\\
1+\sqrt{2} & -1
\end{array}\right].
\label{N101d}
\end{equation}

%------------------------------------------------------------------
% fig.5

\begin{figure}[t!]
\hspace{-3mm} \epsfxsize=8cm \epsfbox{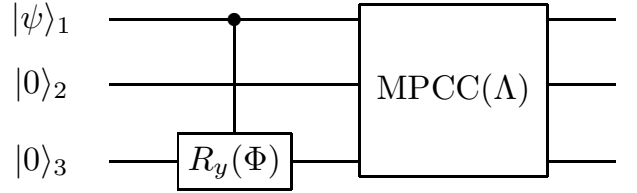} \caption[]{A
quantum circuit, shown in Fig.~\ref{fig5}, implementing the
optimal phase-independent cloning transformation, is equivalent to
the MPCC~\cite{Bartkiewicz} with $\Lambda=\cos\alpha_+$ together
with the controlled rotation  $R_y$ by angle
$\Phi=2(\alpha_--\alpha_+)$.}\label{fig6}
\end{figure}

The optimal phase-independent cloning can be implemented with the
use of a quantum circuit, the MPCC~\cite{Bartkiewicz} setting for
$\Lambda=\cos{\alpha_+}$ (see  Fig.~\ref{fig6}). Therefore, with
minor modifications (concerning the state preparation of the third
qubit), the optimal phase-independent cloning machine can be
realized with, e.g., linear optics or quantum dots as
described for the MPCC in Refs.~\cite{Bartkiewicz,Bartkiewicz2, Miran}.

%------------------------------------------------------------------
% fig.6

%------------------------------------------------------------------
\section{Conclusions}

We analyzed optimal state-dependent cloning of qubit states, which
are described by {\em a priori} known arbitrary phase-independent
(axisymmetric) distribution $g$ on the Bloch sphere. This optimal
cloning reduces in special cases to the universal cloning of
Bu\v{z}ek and Hillery~\cite{Buzek96}, the phase-covariant cloning
of Bru\ss{} {\em et al.}~\cite{Bruss98a} and its generalization by
Fiur\'a\v{s}ek~\cite{F1,F2}, the mirror phase covariant cloning
(MPCC) of Bartkiewicz {\em et al.}~\cite{Bartkiewicz}, or cloning of
an uniform belt of the Bloch sphere of Hu{\em et al.}~\cite{Hu}.

As an example of the state-dependent cloning, we studied the
cloning transformations of qubits described on the Bloch sphere by
the von Mises-Fisher and Brosseau distributions, where the first
is an analog of normal distribution on a sphere~\cite{Fisher} and
the latter describes statistics of the Stokes
parameters~\cite{Barakat, Brosseau}. Whereas the first example is
more abstract and describes Gaussian-like dispersion, the second
example can be used directly to estimate the upper bound for the
capacity of a depolarizing channel~\cite{Cerf00} for photons. Our
results can be also applied in security analysis of various
quantum communication protocols, including quantum
teleportation~\cite{Ozdemir} and quantum key
distribution~\cite{Imre}.

Recently, it was shown that phase independent-cloning can be
parameterized by four parameters~\cite{Wu}. Here, we proved that
only two parameters are sufficient to describe the optimal
phase-independent cloning.

Moreover, we showed that the phase-independent cloning is a simple
generalization of the MPCC and, thus, can be implemented
analogously to the MPCC using photon-polarization
qubits~\cite{Bartkiewicz2} and quantum-dot
spins~\cite{Bartkiewicz}.

{\bf Acknowledgments.} We thank Jaromir Fiur\'a\-\v{s}ek
and Xiaoguang Wang for discussions. A.M. acknowledges support
from the Polish Ministry of Science and Higher Education under
Grant No. 2619/B/H03/2010/38.

%------------------------------------------------------------------

\end{document}